%% file: main.tex
\documentclass[journal]{IEEEtran}
\usepackage[utf8]{inputenc}
\usepackage[T1]{fontenc}
\usepackage{amsmath,amssymb,amsthm,bm}
\usepackage{booktabs}
\usepackage{graphicx}
\usepackage{cite}
\usepackage{tikz}
\usetikzlibrary{positioning,arrows.meta,shapes.geometric,fit,calc}
\usepackage{pgfplots}
\pgfplotsset{compat=1.17}
\definecolor{acc}{RGB}{160,90,23}
\definecolor{ok}{RGB}{46,107,62}
\definecolor{warn}{RGB}{140,106,31}
\definecolor{bad}{RGB}{160,59,46}
\definecolor{muted}{RGB}{110,106,97}
\usepackage[hidelinks]{hyperref}
\usepackage{xcolor}

\newtheorem{proposition}{Proposition}
\newcommand{\pg}{\bm{p}}
\newcommand{\dload}{\bm{d}}
\newcommand{\E}{\mathbb{E}}
\newcommand{\Rset}{\mathcal{R}}
\newcommand{\Uset}{U}
\newcommand{\rev}[1]{#1} %
\colorlet{revcolor}{black} %
\DeclareMathOperator*{\minimise}{minimise}

\begin{document}

\title{When Does Learned Constraint Selection Pay in Security-Constrained Optimisation? A Boundary Map from SCOPF to Unit Commitment}

\author{Milad~Kazemi and
        Behzad~Kazemtabrizi%
\thanks{M.~Kazemi is with the Department of Computer Science, Durham University, Durham, U.K. (e-mail: milad.kazemi@durham.ac.uk).}%
\thanks{B.~Kazemtabrizi is with the Department of Engineering, Durham University, Durham, U.K. (e-mail: behzad.kazemtabrizi@durham.ac.uk).}%
}%

\maketitle

\input{sections/00_abstract}

\input{sections/01_introduction}
\input{sections/02_related}
\input{sections/03_problem}
\input{sections/04_criterion}
\input{sections/05_fixed}
\input{sections/06_reconcile}
\input{sections/07_topology}
\input{sections/08_uc}

\input{sections/09_discussion}
\input{sections/10_conclusion}

\bibliographystyle{IEEEtran}
\bibliography{refs}

\end{document}

%% file: sections/00_abstract.tex
\begin{abstract}
\rev{Learning-to-optimise methods for security-constrained optimal power flow (SCOPF) predict the security constraints that bind on each instance and solve only the reduced problem, with reported speed-ups of two to four orders of magnitude over the full problem. We ask whether learned selection beats the strongest learning-free alternative: a fixed \emph{umbrella} set enforcing every constraint observed to bind in training. Our criterion, the \emph{price of a spare}, answers this in advance: an enforced but inactive constraint is free, so the umbrella is exactly optimal wherever feasible, and learning can pay only where a spare decision carries a price. We test it systematically, including every PGLib-OPF network up to 10\,000 buses. For SCOPF on a fixed network the umbrella is never beaten: its median size is five constraints across $43$ networks, with no size trend, and our selector, while reproducing speed-ups of published magnitude (about $\times80\,000$ at case500), gains at most $0.25$\,ms over the umbrella and violates constraints on $4$--$5\%$ of instances. Configuration diversity can break the umbrella's coverage on a small, densely meshed network but not at 2\,000 buses, and we certify coverage with conformal risk control. In unit commitment, where a spare has a price, a static commitment loses $4$--$12\%$ out of sample, and a predict-repair-reduce pipeline recovers it (cost gap $0.2$--$2.9\%$, fully feasible, $16$--$161\times$ faster than the MILP). In our experiments, whether learning pays depends not on problem size but on whether a spare carries a price.}
\end{abstract}


%% file: sections/01_introduction.tex
\section{Introduction}
\IEEEPARstart{S}{ecurity}-constrained optimal power flow (SCOPF) determines an optimal generation schedule while ensuring that network operating constraints (e.g. branch flow limits) are satisfied under both normal (pre-contingency) and abnormal (post-contingency) operating conditions, in accordance with a pre-defined security criterion such as the N-1 and subject to the permitted system response following a contingency. SCOPF is inherently a much larger problem than its conventional OPF counterpart, mainly due to the number of post-contingency constraints which grows with system size: even in its preventive DC form the total number of constraints could be in the order of millions (about 23 million rows on a 2\,000-bus network), even though most of them are slack and not binding. This has led to a variety of methods for solving a reduced problem that enforces only the constraints needed to return the full problem's optimum. There are two families of methods for identifying these so-called \textit{umbrella} constraints, namely, the \emph{Umbrella constraint discovery} \cite{ardakani2013umbrella} identifies, once for an operating domain, the \textit{minimum} set of constraints that actually enclose the feasible region and enforces only those. \emph{Learning-to-optimise} (L2O) methods \cite{misra2022learning,deka2019learning,le2026hybrid} instead train a model to predict, for each operating instance, the constraints that will bind at that instance's optimum, and solve only the reduced problem over the predicted per-instance binding set. It is important to note that the binding set of constraints is the subset of umbrella constraints that are active at the optimum. Hence, every binding constraint belongs to the umbrella set, whereas constraints in the umbrella set need not be binding at a particular optimum. The L2O methods already report acceleration in solution of two to four orders of magnitude when compared to solving the full SCOPF problem \cite{misra2022learning,deka2019learning,le2026hybrid}. However, the two families of methods are rarely compared to each other.

The L2O literature benchmarks against solving the full problem whereas the umbrella literature focuses on computing the umbrella set efficiently rather than answering whether a per-instance alternative could outperform it. 

This paper compares the two directly, and it does so through a criterion that can be evaluated from the problem's data before any model is trained, so that the outcome is predicted rather than merely observed. The criterion rests on an elementary observation: an enforced constraint that is inactive changes neither the optimiser nor the optimal value. Such a \emph{spare} constraint is therefore free from the perspective of the objective function. Consequently, on any instance whose active set \rev{is contained in} the umbrella, the umbrella is not approximately optimal but exactly optimal, and a per-instance selector can only match it. \rev{Learned selection can therefore pay only through channels other than the objective: the computational cost of carrying spares, or a coverage failure when} the umbrella misses \rev{an active constraint. On a fixed network such failures are rare (none of the $3\,000$ test instances at case118 \rev{or} case500). They become substantial only under network-configuration diversity, on the small, densely meshed case118 (Section~\ref{sec:topology}).} In circumstances in which the main goal of the problem is \rev{to determine} a \emph{decision} (or a set of decisions) that carries \rev{a} cost in the objective, as a committed generating unit does through its no-load cost \rev{in the unit commitment problem}, the observation no longer applies\rev{: a spare committed unit incurs its no-load cost whether or not it is needed, so committing more than an instance requires is no longer free, and a static policy is no longer exactly optimal.} We treat this criterion as a falsifiable prediction and test it systematically.

The main contributions are as follows:

\begin{enumerate}
\item \rev{We state the price-of-a-spare criterion, which predicts in advance that no SCOPF variant (preventive or corrective) admits a regime in which learned selection outperforms the umbrella, whereas unit commitment does (Section~\ref{sec:criterion}).}
\item \rev{We test the SCOPF prediction across six dimensions (grid size, demand distribution, contingency order, corrective redispatch, emergency-rating tightness and the full PGLib-OPF library up to 10\,000 buses) and find no regime in which a per-instance selector outperforms the umbrella. The umbrella, i.e., the union of the per-instance active sets, has a median of five constraints across $43$ networks with no size trend (Pearson $r=0.15$ with bus count), and where it is large the per-instance active set is correspondingly large (e.g., $\E|S|\approx161$ for $|\Uset|=279$ on case4917\_goc), so there is little for a selector to remove (Section~\ref{sec:fixed}).}
\item \rev{We reconcile this with the published L2O speed-ups: our selector reproduces them against the exact solver but loses to the umbrella once re-baselined against it, and accounting for the offline cost of labelling and training does not change this (Section~\ref{sec:reconcile}).}
\item \rev{We identify network-configuration diversity as the one dimension that can break the umbrella's coverage, on a small, densely meshed network, and we give a distribution-free conformal risk control certificate that accepts the large networks and refuses the small one; the effect shrinks with grid size and is absent at 2\,000 buses (Section~\ref{sec:topology}).}
\item \rev{We show that unit commitment is where the criterion flips: a static commitment loses $4$--$12\%$ out of sample, tracking the price of a spare rather than problem size, and a predict-repair-reduce pipeline recovers it at case118 and at nine times the committable fleet (Section~\ref{sec:uc}).}
\end{enumerate}
The remainder of the paper is organised as follows. Section~\ref{sec:related} positions the work. Section~\ref{sec:problem} states the problem and the two policies. Section~\ref{sec:criterion} gives the criterion. Sections~\ref{sec:fixed}--\ref{sec:uc} report the experiments. Section~\ref{sec:discussion} discusses limitations and Section~\ref{sec:conclusion} concludes.

%% file: sections/02_related.tex
\section{Related Work}
\label{sec:related}
The observation that only a small subset of \rev{the full} set of enforced SCOPF constraints\rev{, $\mathcal{C}$,} characterises the potential boundaries of the feasible region for all operating conditions has been known for a long time \cite{stott2012opf}. It follows that the majority of the computational expenditure of SCOPF is due to the large size of the global \rev{constraint set,} most of which do not impact the optimum across all operating domains and hence can be safely discarded. The challenge has since been to propose and implement the most efficient methods to prune and reduce the global \rev{constraint set} to identify only the subset of constraints that potentially shape the boundaries of the feasible region. To that end, Ardakani and Bouffard \cite{ardakani2013umbrella} proposed a precise mathematical formulation for discovering what they defined as the \textit{\rev{umbrella}} set, \rev{the minimal set of constraints that shape the feasible region, i.e., the constraints whose removal would change the set of feasible solutions. This set is geometric: it depends on the constraints but not on the objective. In this paper we use the term for an objective-dependent subset of it, namely the union of optimal active sets over the operating domain, $U(\Theta)=\bigcup_{\theta\in\Theta}S(\theta)$,} where $\Theta$ represents the operating domain of interest as well as other parameters of interest, for example network \rev{configuration}, and $S(\theta)\rev{\subseteq\mathcal{C}}$ \rev{denotes} the set of constraints that are binding at the optimal solution for a particular operating condition $\theta$\rev{, i.e.,} the active set. Thus, $U\rev{(\Theta)}$ comprises all constraints that can potentially become active as the operating condition varies over $\Theta$\rev{. Since a constraint that binds at some optimum lies on the boundary of the feasible region, $U(\Theta)$ is contained in the umbrella set of \cite{ardakani2013umbrella} (up to weakly redundant constraints, such as duplicated rows), and it can be strictly smaller, because not every boundary constraint binds at an optimum for a given cost function.} \rev{Ardakani} and Bouffard \rev{first developed} a model to discover the \rev{umbrella} set for DC-based SCOPF problems, and later proposed \rev{accelerated} \cite{ardakani2015acceleration} and \rev{learning-based} \cite{ardakani2018prediction} \rev{variants of} the same discovery procedure. In all cases, the objective was to identify a reduced set of constraints and solve a reduced optimisation problem with only those constraints enforced. Our static baseline is a sampling-based estimate of \rev{$U(\Theta)$,} and as such we adopt the same name and claim no novelty for it. The novelty of our work lies not \rev{in} how to compute the umbrella set \rev{but in} whether a per-instance alternative could \rev{outperform} it. 

Learning active sets was formalised by Misra \emph{et al.} \cite{misra2022learning}, who observed that few distinct active sets occur over a parameter distribution. Ng \emph{et al.} \cite{ng2018statistical} and Deka and Misra \cite{deka2019learning} classify DC-OPF active sets with neural networks\rev{, and Robson \emph{et al.} \cite{robson2019learning} learn a reduced OPF formulation through meta-optimisation}. Our predictor belongs to this family, and our results sharpen rather than contradict theirs in that a few distinct active sets is precisely the condition under which the union of those sets is small, hence the condition under which enforcing \rev{the umbrella} outperforms predicting each per-instance set. Constraint screening \cite{pineda2020data,porras2023cost,wu2024efficient,yang2023costaware} removes provably redundant constraints. In \cite{porras2023cost}, Porras \emph{et al.} note that learned elimination may be over-optimistic and produce infeasible solutions\rev{. This is} the failure mode we measure for learned selectors\rev{ (Section~\ref{sec:fixed}),} and the one our \rev{coverage} certificate bounds \rev{ for the umbrella, by bounding the probability that a new configuration binds a constraint outside the enforced set (Section~\ref{sec:topology})}. Closest in spirit \rev{ to our work} is Pineda and Morales \cite{pineda2022lowhanging}, who show that a \rev{simple nearest-neighbour} method for unit commitment\rev{, which fixes the commitments of the most similar past instances and keeps the cheapest feasible dispatch among them,} is difficult to beat\rev{. We implement their method as a baseline (Section~\ref{sec:uc}): it is fully feasible and near-optimal in single-period unit commitment, but leaves multi-period days that no historical commitment can serve. We offer an explanation of when such a simple baseline is strong, and why.} On the method side we build on the hybrid mixed-integer L2O framework of Le \emph{et al.} \cite{le2026hybrid} with a differentiable QP layer \cite{amos2017optnet}\rev{. As a baseline we include direct prediction of the dispatch, the simplest form of the optimisation} proxies \rev{(surrogate models that predict the solution directly) studied in} \cite{donti2021dc3,zhao2023ensuring}\rev{, here without their feasibility-restoration steps}. Certification of learned optimisation has followed worst-case verification \cite{venzke2020worstcase,nellikkath2022pinn,chen2024scalable} and distribution-free calibration \cite{vovk2005algorithmic,angelopoulos2023gentle,angelopoulos2024crc,angelopoulos2021learn}, recently applied to feasibility \cite{patel2025conformal} and optimality gaps \cite{zhang2025tightening}. We add a coverage guarantee on the enforced constraint set itself\rev{, which bounds the risk that the reduced problem returns a solution that is infeasible for the full problem (Section~\ref{sec:topology})}.

%% file: sections/03_problem.tex
\section{Problem Statement}
\label{sec:problem}

\subsection{Preventive DC-SCOPF}
Let $\pg\in\mathbb{R}^{n_g}$ be generator outputs, $\dload$ the nodal load, $f_l(\pg,\dload)$ the DC flow on branch $l$, $\Psi_{l,k}$ the line outage distribution factor of branch $l$ with respect to outage $k$, and $\mathcal{K}$ the set of non-islanding single-branch outages. Preventive DC-SCOPF is
\begin{subequations}\label{eq:scopf}
\begin{align}
\minimise_{\pg}\quad & \textstyle\sum_g c_{2,g}p_g^2 + c_{1,g}p_g \\
\text{s.t.}\quad & \bm 1^{\!\top}\pg = \bm 1^{\!\top}\dload,\qquad p_g^{\min}\le p_g\le p_g^{\max},\\
& |f_l(\pg,\dload)|\le r^A_l,\quad l=1,\dots,n_l,\\
& |f_l(\pg,\dload)+\Psi_{l,k}f_k(\pg,\dload)|\le r^C_l,\ \ l\neq k,\ k\in\mathcal{K}. \label{eq:cont}
\end{align}
\end{subequations}
Since $f=A\pg+f^{\mathrm{load}}(\dload)$ with $A=\mathrm{PTDF}\,C_g$, all flow limits stack into one sparse system $F\pg\le h(\dload)$ in which only the right-hand side varies per instance. \rev{Each row of this system is one base-case or post-contingency flow constraint.} Rows that provably cannot bind for any load in the sample are removed by a safe box screen (row $r$ is redundant if $\max_{p^{\min}\le\pg\le p^{\max}}(F\pg)_r\le\min_i h_r^i$), which never removes a binding row. \rev{In PGLib-OPF the emergency rating equals the normal rating,} $r^C=r^A$, under which preventive N-1 is infeasible \rev{for almost all instances of the networks in Fig.~\ref{fig:tightness}}. Following the convention of the L2O literature we use an emergency-rating uplift $r^C=1.5\,r^A$ unless stated otherwise, and Section~\ref{sec:fixed} sweeps it. A small quadratic regulariser makes the linear-cost cases strictly convex, so the optimiser is unique (a well-defined learning target). The induced cost distortion is about $0.05\%$ and labels and the learned model use the same regularised problem.

\subsection{Two policies}
Let $\theta$ be the instance parameter (the load, and later the network configuration), $\Rset$ the screened candidate set, $S(\theta)\subseteq\Rset$ the constraints active at the optimum, and $\Uset=\bigcup_{\theta\in\mathcal{T}}S(\theta)$ the \emph{umbrella}, the union of active sets over a training sample $\mathcal{T}$\rev{, which estimates $U(\Theta)$ of Section~\ref{sec:related}}. Write $\mathrm{OPT}(\theta)$ for the true optimal value, $v(\mathcal{E},\theta)$ for the value of the reduced problem enforcing only $\mathcal{E}\subseteq\Rset$, and $T(k)$ for the wall-clock cost of a reduced solve with $k$ enforced \rev{constraints}. The \emph{umbrella policy} enforces the fixed set $\Uset$ on every instance and uses no learned parameters. A \emph{selector} predicts $\widehat S(\theta)$ per instance and enforces that.

\subsection{The learned selector, and the umbrella}
Our selector follows the hybrid L2O architecture of \cite{le2026hybrid} adapted to SCOPF (Fig.~\ref{fig:pipeline}). A multilayer perceptron $\pi_\omega(\dload)$ (three hidden layers of $256$ ReLU units) maps the load to logits $\bm\ell$ over \rev{a candidate set of $K$ screened constraints: the training-set umbrella plus the $16$ tightest constraints outside it, with near-parallel duplicates removed ($K=17$ of $3\,637$ at case118).}  Next, a straight-through estimator \rev{\cite{bengio2013estimating}} produces the binary selection $\bm\delta$, and a differentiable reduced QP recovers the dispatch $\pg$ under the selected \rev{constraints}. Consequently, the training loss \rev{\eqref{eq:loss}} combines the recovered cost, a self-supervised \rev{penalty on the violation of} \emph{all} screened \rev{constraints by the recovered dispatch, in proportion to the overload, with a stop-gradient complementarity correction that charges this violation to the deselected candidates, so} that deselecting a needed \rev{constraint, i.e., one which the dispatch then violates,} is penalised, and a recall-weighted binary cross-entropy against the label\rev{led} active set\rev{:}
\begin{equation}\label{eq:loss}\color{revcolor}
\begin{aligned}
\mathcal{L}={}&\gamma_{\mathrm o}\,c(\pg)/\bar c+\gamma_{\mathrm s}\,\mathrm{BCE}_w(\bm\ell,\bm\delta^\star)\\
&+\gamma_{\mathrm v}\textstyle\sum_{r\in\Rset}[F_r\pg-h_r]_+\\
&+\gamma_{\mathrm v}\textstyle\sum_{k=1}^{K}\bigl(1-\sigma(\ell_k)\bigr)\,\mathrm{sg}\bigl([F_k\pg-h_k]_+\bigr),
\end{aligned}
\end{equation}
\rev{where $F_r\pg\le h_r$ is constraint $r$, $c$ is the generation cost, $\bar c$ the mean training-label cost, $[\cdot]_+=\max(\cdot,0)$, $\sigma$ the logistic function, $\mathrm{sg}$ the stop-gradient, $\bm\delta^\star$ the labelled active set and $w=2$ the positive-class weight, with $\gamma_{\mathrm o}=1$, $\gamma_{\mathrm v}=10$ and $\gamma_{\mathrm s}=1$. The QP layer and the labels both add the regulariser $\|\pg\|^2$ (in p.u.) to $c$.} Ablations that drop the supervised term (SSL), drop the self-supervised term (SL-active-set), or regress the dispatch directly (SL-direct) are reported\rev{; Table~\ref{tab:models} summarises how each model is trained and deployed}. \rev{Deselected candidates are released by adding $M(1-\delta_k)$ to their limit, with $M=100$.} Two gradient corrections, \rev{the stop-gradient term in \eqref{eq:loss} and a surrogate gradient for this gate that treats $M$ as $1$ in the backward pass}, are needed for the selection head to receive a usable \rev{gradient: without the first, a released constraint has exactly zero gradient, and without the second, the gate's gradient, scaled by $M$, swamps the cross-entropy term}. At deployment every \rev{QP-layer} mode is followed by a restoration loop that adds any violated \rev{constraints} and re-solves however, \rev{SL-direct has no QP and is used as predicted}. The umbrella policy uses the same reduced QP with the fixed set $\Uset$ and no learned selector.

\begin{table}[t]\color{revcolor}
\caption{Models compared. All learned modes share the same MLP.}
\label{tab:models}
\centering\scriptsize\setlength{\tabcolsep}{3pt}
\begin{tabular}{lccc}
\toprule
Model & trained on labelled & trained on cost and & reduced QP at \\
      & active set (BCE) & violation (through QP) & deployment \\
\midrule
Hybrid (full model) & yes & yes & yes \\
SSL & no & yes & yes \\
SL-active-set & yes & no & yes \\
SL-direct & regresses $\pg$ directly & no & no \\
Umbrella & no training; fixed $\Uset$ & n/a & yes \\
\bottomrule
\end{tabular}
\end{table}

\input{figures/fig_pipeline}

%% file: figures/fig_pipeline.tex
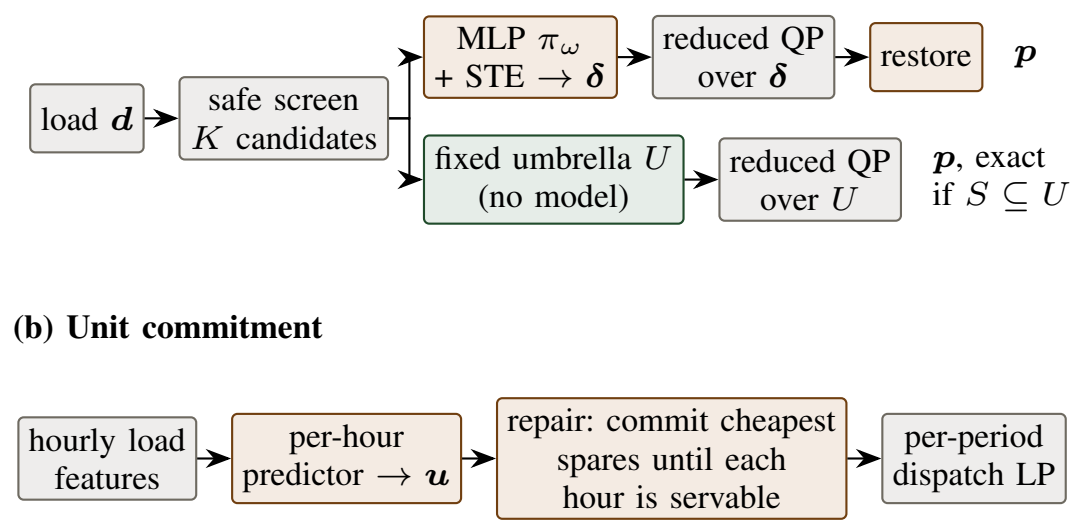
\begin{figure}[t]\centering
\begin{tikzpicture}[scale=0.9, transform shape, font=\scriptsize, >=Stealth, node distance=2.5mm,
  box/.style={draw, rounded corners=1pt, align=center, inner sep=2.2pt, minimum height=5mm},
  lrn/.style={box, fill=acc!12, draw=acc!70!black},
  stc/.style={box, fill=ok!12, draw=ok!70!black},
  fix/.style={box, fill=muted!12, draw=muted},
  lab/.style={font=\scriptsize\bfseries, anchor=west, inner sep=0pt}]
  \node[lab] at (0,1.05) {(a) SCOPF};
  \node[fix] (load) at (0.55,0) {load $\dload$};
  \node[fix, right=of load] (scr) {safe screen\\$K$ \rev{candidates}};
  \node[lrn, right=of scr, yshift=4.5mm] (mlp) {MLP $\pi_\omega$\\+ STE $\to\bm\delta$};
  \node[stc, right=of scr, yshift=-4.5mm] (umb) {fixed \rev{umbrella} $\Uset$\\(no model)};
  \node[fix, right=of mlp] (qp1) {reduced QP\\over $\bm\delta$};
  \node[fix, right=of umb] (qp2) {reduced QP\\over $\Uset$};
  \node[lrn, right=of qp1] (res) {restore};
  \draw[->] (load) -- (scr);
  \draw[->] (scr.east) -- ++(1.5mm,0) |- (mlp.west);
  \draw[->] (scr.east) -- ++(1.5mm,0) |- (umb.west);
  \draw[->] (mlp) -- (qp1); \draw[->] (umb) -- (qp2); \draw[->] (qp1) -- (res);
  \node[right=1mm of res] {$\pg$};
  \node[right=1mm of qp2, align=left] {$\pg$, exact\\if $S\subseteq\Uset$};
  \node[lab] at (0,-1.55) {(b) Unit commitment};
  \node[fix] (ld) at (0.7,-2.5) {hourly load\\features};
  \node[lrn, right=of ld] (pred) {per-hour\\predictor $\to\bm u$};
  \node[lrn, right=of pred] (rep) {repair: commit cheapest\\spares until each\\hour is servable};
  \node[fix, right=of rep] (lp) {per-period\\dispatch LP};
  \node[right=1mm of lp] {schedule};
  \draw[->] (ld) -- (pred); \draw[->] (pred) -- (rep); \draw[->] (rep) -- (lp);
\end{tikzpicture}
\caption{The two pipelines. (a) SCOPF: the learned selector (shaded) and the umbrella share the screen and the reduced QP; they differ only in who chooses the enforced \rev{constraints}. (b) Unit commitment: predict, repair, reduced solve; no integer variables at deployment.}
\label{fig:pipeline}
\end{figure}

%% file: sections/04_criterion.tex
\section{\rev{Free Spares and} the Price of a Spare}
\label{sec:criterion}
\begin{proposition}[Spare invariance]\label{prop:spare}
Consider a convex programme with strictly convex objective and a constraint qualification, with unique optimiser $z^\star$ and active set $S$. If $S\subseteq\mathcal{E}\subseteq\Rset$, the reduced problem enforcing only $\mathcal{E}$ has the same optimal value and the same optimiser: $v(\mathcal{E})=\mathrm{OPT}$ and $z_{\mathcal{E}}=z^\star$.
\end{proposition}
\begin{proof}
At $z^\star$ the KKT conditions of the full problem hold with multipliers supported on $S$. The dropped constraints $\Rset\setminus\mathcal{E}$ are inactive at $z^\star$, so the same multipliers certify the KKT conditions of the reduced problem, which are sufficient by convexity; hence $z^\star$ is optimal for the reduced problem, and by strict convexity it is its unique optimiser.
\end{proof}

It follows that strict convexity matters for the purposes of this paper. For a linear-cost SCOPF the reduced problem could return a different optimiser of equal value that violates a dropped \rev{constraint}, which is why labels and every solver in this paper use the regularised, strictly convex problem. Proposition~\ref{prop:spare} says that an enforced but inactive constraint is free\rev{, which we call the \emph{free-spare property}:} on any instance that the umbrella covers, it is exactly optimal, and a learned selector can \rev{at best} match it on \rev{objective} cost. The added value of the learned \rev{selector} should therefore come from \rev{three} channels\rev{,} namely \textbf{(C1) Computational}, \textbf{(C2) Objective}, and \textbf{(C3) Coverage}. \textbf{(C1)} \rev{states} that carrying $|\Uset|-\E|S|$ spares \rev{costs} $T(|\Uset|)-T(\E|S|)$ per query; \textbf{(C2)} \rev{states} that \rev{where} the discrete object is a \emph{decision} that carries cost, such as a committed unit incurring a no-load cost \rev{in unit commitment}, Proposition~\ref{prop:spare} does not apply and a spare costs money on every instance that does not need it; and \textbf{(C3)} states that if $S(\theta)\not\subseteq\Uset$\rev{,} the umbrella\rev{'s reduced solution is infeasible for the full problem and must be restored at a cost $\kappa$}. Writing $\pi$ for the price of a spare through whichever of (C1)--(C2) is active, and $p^{\mathrm U}_{\mathrm{miss}}\rev{=\Pr[S(\theta)\not\subseteq\Uset]}$, $p^{\mathrm{sel}}_{\mathrm{miss}}\rev{=\Pr[S(\theta)\not\subseteq\widehat S(\theta)]}$ for the probabilities \rev{that the umbrella and the selector, respectively, omit an} active constraint, the expected advantage of a selector over the umbrella is
\begin{equation}\label{eq:criterion}
\mathcal{A}=\pi\bigl(|\Uset|-\E|S(\theta)|\bigr)+\bigl(p^{\mathrm U}_{\mathrm{miss}}-p^{\mathrm{sel}}_{\mathrm{miss}}\bigr)\kappa ,
\end{equation}
It follows that learned selection is only worth building if $\mathcal{A}>0$\rev{, and we call a setting with $\mathcal{A}>0$ a \emph{regime} for learned selection}. For a perfect selector $p^{\mathrm{sel}}_{\mathrm{miss}}=0$, so \eqref{eq:criterion} with that substitution upper-bounds what \emph{any} trained model can deliver, and it can be measured without training. It is convenient to summarise the geometry by the \emph{learnability index} $\rho=\E|S(\theta)|/|\Uset|\in(0,1]$: $\rho\to1$ means instances bind essentially the whole umbrella and there is nothing to strip, whereas $\rho\ll1$ is necessary but not sufficient for selection to pay\rev{, i.e., for a selector to outperform the umbrella through any of the three channels}.

Fig.~\ref{fig:map} places the settings of this paper on the two axes of \eqref{eq:criterion}. In any SCOPF variant the discrete object is a set of constraints, so only (C1) is available and $T$ is flat over the relevant range (we measure $T(9)=\rev{1.1}$\,ms against a $\rev{286}$\,ms exact solve at case118); scale, contingency order, load spread and corrective redispatch change $|\Uset|$ and $\rho$ but not $\pi$, so none of them should open a regime. Where a spare carries an objective price, in unit commitment, $\mathcal{A}>0$ and should grow with that price rather than with fleet size. And if configuration diversity makes $\Uset$ non-enumerable, $p^{\mathrm U}_{\mathrm{miss}}$ becomes positive and coverage must be certified rather than assumed.

\input{figures/fig_map}

%% file: figures/fig_map.tex
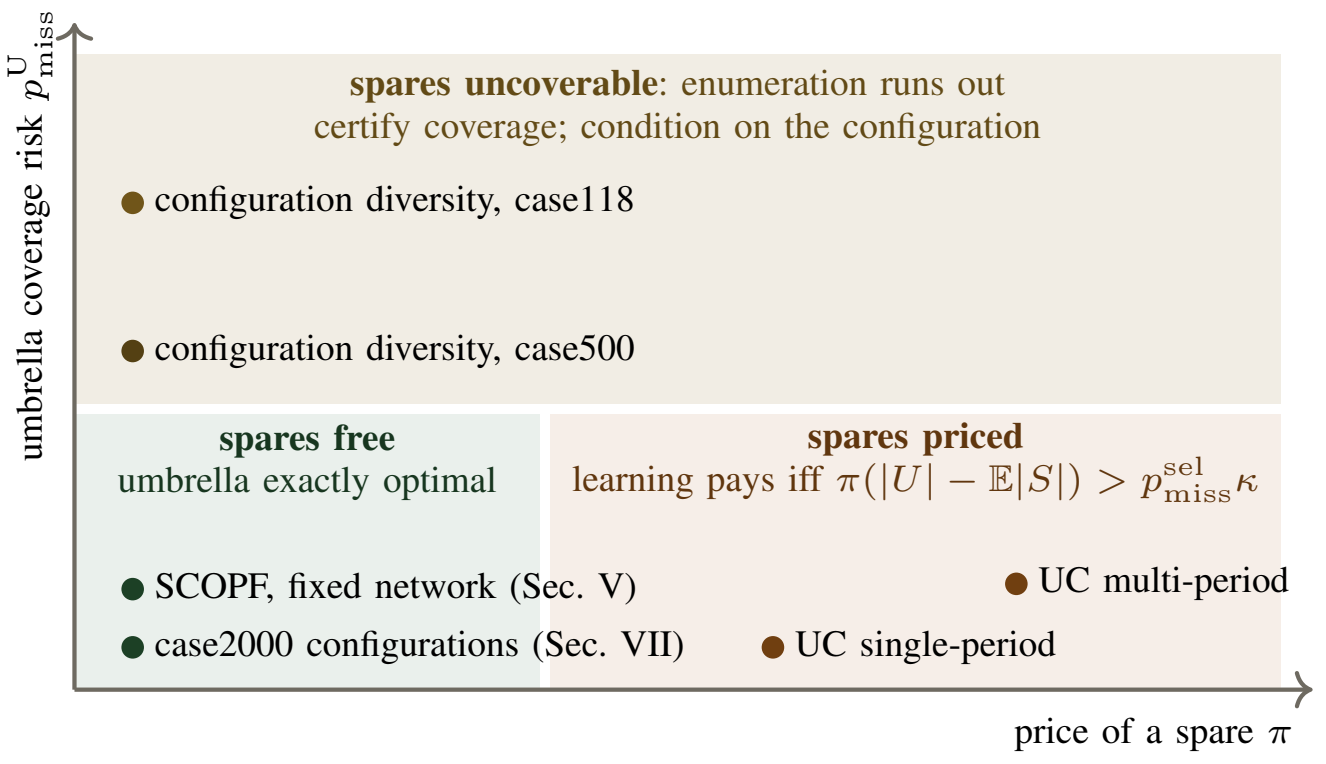
\begin{figure}[t]\centering
\begin{tikzpicture}[scale=0.70, font=\scriptsize, pt/.style={circle, inner sep=1.5pt}, plab/.style={font=\scriptsize, anchor=west, inner sep=1pt}]
  \fill[ok!10]   (0,0)   rectangle (4.4,2.6);
  \fill[acc!10]  (4.5,0) rectangle (11.4,2.6);
  \fill[warn!12] (0,2.7) rectangle (11.4,6.0);
  \draw[->, thick, muted] (0,0) -- (11.7,0);
  \draw[->, thick, muted] (0,0) -- (0,6.3);
  \node at (10.2,-0.42) {price of a spare $\pi$};
  \node[rotate=90] at (-0.42,4.3) {umbrella coverage risk $p^{\mathrm U}_{\mathrm{miss}}$};
  \node[ok!50!black, align=center] at (2.2,2.15) {\textbf{spares free}\\ umbrella exactly optimal};
  \node[acc!60!black, align=center] at (7.95,2.15) {\textbf{spares priced}\\ learning pays iff $\pi(|U|-\mathbb E|S|)>p^{\mathrm{sel}}_{\mathrm{miss}}\kappa$};
  \node[warn!70!black, align=center] at (5.7,5.5) {\textbf{spares uncoverable}: enumeration runs out\\ certify coverage; condition on the configuration};
  \node[pt, fill=ok!60!black] at (0.55,0.95) {}; \node[plab] at (0.7,0.95) {SCOPF, fixed network (Sec.~V)};
  \node[pt, fill=ok!60!black] at (0.55,0.4) {}; \node[plab] at (0.7,0.4) {case2000 configurations (Sec.~VII)};
  \node[pt, fill=acc!70!black] at (6.6,0.4) {}; \node[plab] at (6.75,0.4) {UC single-period};
  \node[pt, fill=acc!70!black] at (8.9,1.0) {}; \node[plab] at (9.05,1.0) {UC multi-period};
  \node[pt, fill=warn!80!black] at (0.55,4.6) {}; \node[plab] at (0.7,4.6) {configuration diversity, case118};
  \node[pt, fill=warn!60!black] at (0.55,3.2) {}; \node[plab] at (0.7,3.2) {configuration diversity, case500};
\end{tikzpicture}
\caption{The criterion as a map. Fixed-network SCOPF sits at the origin (free spares, closed coverage), whatever $\rho$ looks like. Unit commitment moves right and learning pays. Configuration diversity moves up on a small dense network, where coverage must be certified, and returns to the origin at $2\,000$ buses.}
\label{fig:map}
\end{figure}

%% file: sections/05_fixed.tex
\section{Fixed Network: \rev{No} Regime \rev{for Learned Selection}}
\label{sec:fixed}

\subsection{Setup}
We have set up fixed-network experiments using the demand data in OPFData \cite{opfdata2024} for pglib case118 and case500\_goc ($15\,000$ instances each; $\pm 20\%$ per-bus perturbation), with a $\pm40\%$ widened envelope as a stressed variant. Labels are exact solves (i.e., exact solutions of the DC-SCOPF for each instance per network) with Clarabel \cite{clarabel2024}\rev{ over the full screened constraint set; for case2000, the larger library networks and the configuration study (Section~\ref{sec:topology}) they are computed} through a lazy working-set solver\rev{, i.e., constraint generation that starts from the base-case constraints, adds every violated screened constraint and re-solves until none remains, which we validated against the full solve on case118}. The library sweep uses PGLib-OPF \cite{pglib2019} with demands resampled at the same width. \rev{Learned} selectors are trained on $10\,000$ labels and evaluated on $3\,000$ held-out instances.

\subsection{Head-to-head at case118}
Table~\ref{tab:h2h} gives the comparison for each trained model for the case118 system as an example. Every QP-layer mode is near-exact \rev{in cost} whenever \rev{its selected set contains the true active set (Proposition~\ref{prop:spare})}, so the \rev{models in Table~\ref{tab:h2h} differ mainly} in their feasibility \rev{(violation rate and worst violation)} and \rev{in the} structure \rev{of the selected set (precision and recall against the true active set)}. Restoration repairs every \rev{QP-layer} mode to full feasibility at no measurable cost in $0.04$ mean iterations, however the umbrella never requires any restoration. Across five seeds the hybrid (full model) is stable ($4.00\pm\rev{1.05}\%$ violations\rev{, mean and sample standard deviation}) whilst pure self-supervision (\rev{trained without labels}) is not (per-seed violation $0.0$, $1.6$, $4.4$, $7.7$, $23.0\%$). An eight-point loss-weight sweep \rev{(not tabulated)} shows a robust plateau with two cliffs: \rev{cutting the supervised weight tenfold} collapses feasibility to $32.6\%$ violations, \rev{and cutting the violation weight tenfold} admits a $1.48$\,p.u.\ worst case. A constant majority-\rev{constraint} selector violates on $35.9\%$ of instances against the hybrid's $4.1\%$, so the network does \rev{learn to} predict \rev{the rare, instance-specific constraints} however, predicting them simply \rev{brings no benefit over enforcing them all. Self-supervision shows the same mechanism from the training side: SSL attains a recall of $0.993$ but a precision of only $0.329$. A spare constraint is free (Proposition~\ref{prop:spare}), so neither the cost nor the violation term gives it a reason to deselect one, and without labelled active sets it over-selects. The larger reduced problem also makes it the slowest QP-layer mode ($\times195$ against $\times241$ for the hybrid).}

\input{tables/tab_h2h}

\subsection{Mechanism: $T(k)$ and $|U(n)|$}
Fig.~\ref{fig:tk} illustrates \rev{the two facts that settle the fixed-network case. First, the} umbrella's only cost channel is $T(k)$, which is superlinear \rev{but} remains at a few milliseconds for every \rev{umbrella} ever observed. \rev{Second,} on a fixed network the \rev{umbrella's} cardinality saturates within the label budget (case118: nine \rev{constraints} by $n=2\,000$, \rev{unchanged over} the next $13\,000$ instances)\rev{, so coverage is not the limiting factor: the umbrella covered every held-out instance we observed}. \rev{The sample this needs depends on the network: on case118, $2\,000$ random training instances cover every test instance in $99\%$ of draws, but at case500 and under the $\pm40\%$ envelope the rarest active constraints bind on only two and one of $10\,000$ training instances, so exactness there requires a large sample.}

\input{figures/fig_tk}

\subsection{Six falsification attempts}
Table~\ref{tab:falsify} summarises \rev{the six} \textit{levers} \rev{(dimensions of the problem that we varied)} we pulled \rev{in an attempt} to open a regime \rev{for learned selection, i.e., a large umbrella of which each instance binds only a small fraction (large $|\Uset|$ and small $\rho$); none of them did}. \rev{On case118,} full N-2 multiplies \rev{the number of} candidates $90\times$ (to $327\,010$ screened \rev{constraints}) yet \rev{$|\Uset|$ grows only} from $9$ to about $13$ and saturates within thirty instances: more contingencies mostly re-stress the same monitored transmission corridors rather than binding new ones. Corrective SCOPF introduces the first structure that scales, ramp constraints whose \rev{umbrella} grows with the \rev{network} ($8\to59\to161$), but at every ramp budget that saves cost the \rev{umbrella} equals the per-instance active set, and at the one tight-budget point with a large \rev{umbrella} the hardest instances bind $146$ of $161$. \rev{Across} $15$ congested PGLib networks the median of $\max|S|/|\Uset|$ is $1.00$. A safe selector must cover the per-instance maximum, so it can strip almost nothing.

The last lever is the emergency rating $r^C$ itself: perhaps between ``loose'' and ``infeasible'' lies a window in which the \rev{umbrella} explodes while instances need only slivers of it. Sweeping the uplift finely from $1.5$ \rev{towards} the infeasibility boundary at fixed load (Fig.~\ref{fig:tightness}) says no, twice, for different reasons. On case118 the window is one rung wide ($1.4\to1.2$), $|\Uset|$ moves $8\to12$ and $\rho\in[0.71,0.80]$ before feasibility collapses. On case500 the window is wider and $\rho$ does dip, to $0.33$ at uplift $1.15$, but $|\Uset|$ never exceeds $12$ \rev{constraints} on the whole axis, so what selection could strip is the difference between enforcing about ten \rev{constraints} and about three. The regime learned selection needs, large $|\Uset|$ \emph{and} small $\rho$ \emph{and} nonzero feasibility, is a conjunction, and each network falsifies a different conjunct before the boundary. Setting $r^C\to0$ is not an option: post-contingency flows are linear images of nonzero injections, so the feasible set empties.

\input{figures/fig_tightness}

\input{tables/tab_falsify}

\subsection{The PGLib library sweep}
To upgrade ``the \rev{umbrella} is small'' from an observation about hand-picked grids to a property of the benchmark library, we measured every PGLib-OPF network up to $10\,000$ buses with a validated MATPOWER parser and the identical pipeline: loads resampled per bus at $\pm20\%$, and each network pushed to the \emph{largest} load scale it can serve securely, i.e.\ its most congested feasible operating point, where the \rev{umbrella} is largest and the claim is hardest to defend (Fig.~\ref{fig:library}). Across $46$ of the $47$ networks, $\rev{43}$ are measurable\rev{; ten of them first appeared infeasible only because our ascending search over load stopped at the first infeasible scale, although they are feasible at higher load (for nine of them the total minimum generation exceeds the demand at half load, and for case4661\_sdet relaxing thermal limits removes the infeasibility). Of the other three, case2853\_sdet is N-1 infeasible at every tested (rating, load) pair, case1803\_snem has a singular DC susceptance matrix (two zero-reactance branches), and case10000\_goc is N-1 infeasible at the base uplift, with higher uplifts untested; we report them} rather than drop \rev{them}. The \rev{umbrella} has \rev{a} median \rev{of} $5$ \rev{constraints}, ninth decile $\rev{38}$ and maximum $\rev{279}$, against up to $48$ million screened \rev{constraints} per network, with no size trend (Pearson $r=\rev{0.15}$ with bus count): the $9\,241$-bus network has $|\Uset|=25$, the $4\,601$-bus one $|\Uset|=1$, and a $9\,591$-bus one $|\Uset|=0$. Where the \rev{umbrella} is large\rev{,} the overlap is high: case4917\_goc at $\rev{279}$ has $\E|S|\approx161$ ($\rho=\rev{0.58}$). The sweep cross-validates against the OPFData pipeline (case118: $|\Uset|=8$ vs.\ $9$, $\E|S|=6.21$ vs.\ $\rev{6.17}$; case500: \rev{$|\Uset|=5$ vs.\ $6$, and} $\E|S|$ agrees to three figures). Two caveats travel with these numbers: the stressed operating points pin active sets (several networks show $\rho=1.00$), so the $\rho$ values describe the boundary of the operating set rather than nominal operation; and $|\Uset|$ is not comparable across networks that required different uplifts, which are reported per network. One network (case8387\_pegase) is a compute footnote rather than a counter example: screening is tractable, but exact labelling of $200$ instances exceeded two 48-hour allocations.

\input{figures/fig_library}

%% file: tables/tab_h2h.tex
\begin{table}[t]
\caption{Case118, $3\,000$ held-out instances. \rev{Viol.\ rate: share of test instances whose reduced solution, before restoration, violates a screened constraint, the power balance or a generator limit by more than $10^{-5}$\,p.u.; P/R: precision and recall of the selected set against the true active set.} Speed-up \rev{is the median exact solve ($286$\,ms) over the median per-query time, i.e., the network's single-query latency plus the reduced solve (learned modes with a QP $1.2$--$1.5$\,ms, SL-direct $0.05$\,ms, umbrella with $\Uset$ $1.2$\,ms), all timed single-threaded on exclusive nodes of one cluster.}}
\label{tab:h2h}
\centering\small
\setlength{\tabcolsep}{2.5pt}\scriptsize
\begin{tabular}{lcccc}
\toprule
Model & viol.\ rate & worst (p.u.) & P / R & speed-up \\
\midrule
Hybrid & $4.10\%$ & $0.161$ & $0.930$ / $0.991$ & \rev{$\times241$} \\
SSL & $3.20\%$ & $0.296$ & $0.329$ / $0.993$ & \rev{$\times195$} \\
SL-active-set & $2.73\%$ & $0.161$ & $0.988$ / $0.994$ & \rev{$\times243$} \\
SL-direct & $100\%$ & $0.454$ & --- & \rev{$\times5\,560$} \\
\textbf{Umbrella} & $\bm{0.00\%}$ & $\sim0$ & --- & \rev{$\times235$} \\
\bottomrule
\end{tabular}
\end{table}

%% file: figures/fig_tk.tex
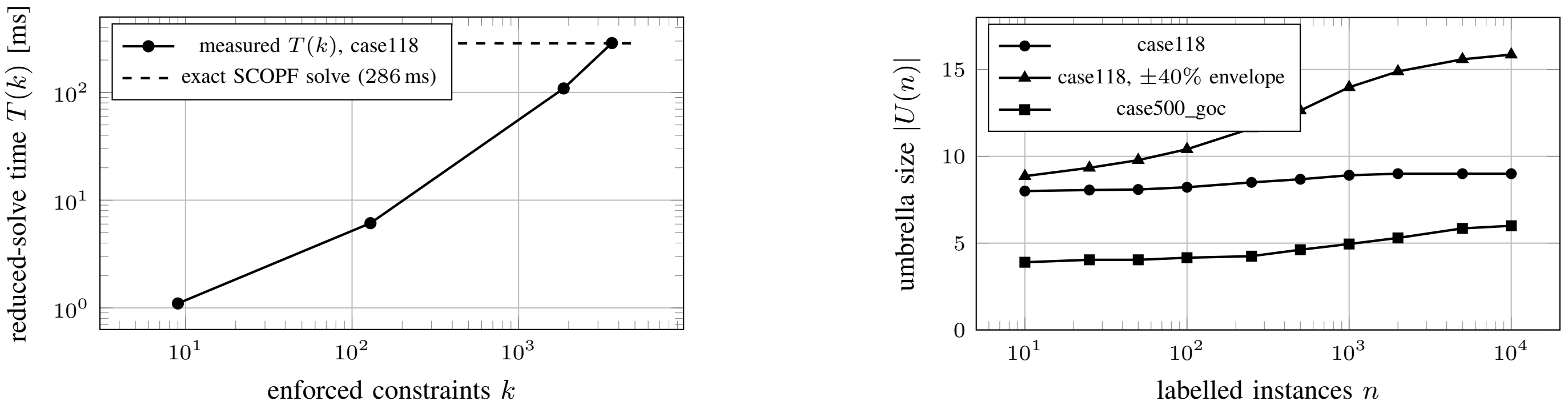
\begin{figure*}[t]\centering
\begin{tikzpicture}
\begin{axis}[width=0.46\textwidth, height=5.2cm, xmode=log, ymode=log,
  xlabel={enforced constraints $k$}, ylabel={reduced-solve time $T(k)$ [ms]},
  tick label style={font=\scriptsize}, label style={font=\small},
  legend style={font=\scriptsize, at={(0.02,0.98)}, anchor=north west}, grid=major]
\addplot[thick, mark=*, mark size=1.6pt, revcolor] coordinates {(9,1.10) (129,6.12) (1866,109.0) (3637,287.6)};
\addlegendentry{measured $T(k)$, case118}
\addplot[dashed, thick, revcolor] coordinates {(6,286) (5000,286)};
\addlegendentry{exact SCOPF solve ($\rev{286}$\,ms)}
\end{axis}
\end{tikzpicture}\hfill
\begin{tikzpicture}
\begin{axis}[width=0.46\textwidth, height=5.2cm, xmode=log,
  xlabel={labelled instances $n$}, ylabel={\rev{umbrella size} $|U(n)|$}, ymin=0, ymax=18,
  tick label style={font=\scriptsize}, label style={font=\small},
  legend style={font=\scriptsize, at={(0.02,0.98)}, anchor=north west}, grid=major]
\addplot[thick, mark=*, mark size=1.4pt] coordinates {(10,8.0) (25,8.06) (50,8.09) (100,8.22) (250,8.5) (500,8.68) (1000,8.91) (2000,9) (5000,9) (10000,9)};
\addlegendentry{case118}
\addplot[thick, mark=triangle*, mark size=1.7pt] coordinates {(10,8.86) (25,9.34) (50,9.78) (100,10.41) (250,11.61) (500,12.64) (1000,13.98) (2000,14.89) (5000,15.59) (10000,15.86)};
\addlegendentry{case118, $\pm40\%$ envelope}
\addplot[thick, mark=square*, mark size=1.4pt] coordinates {(10,3.9) (25,4.04) (50,4.04) (100,4.16) (250,4.25) (500,4.62) (1000,4.95) (2000,5.3) (5000,5.85) (10000,6.0)};
\addlegendentry{case500\_goc}
\end{axis}
\end{tikzpicture}
\caption{Left: reduced-solve time against the number of enforced \rev{constraints (median over the same test instances, one compiled solver per $k$, single-threaded on one node; dashed: the exact solve timed the same way)}; at every \rev{umbrella} observed on a fixed network ($6$--$19$ \rev{constraints}) it is a few milliseconds. Right: \rev{mean umbrella size over random training subsets of size $n$}; it saturates within the label budget.}
\label{fig:tk}
\end{figure*}

%% file: figures/fig_tightness.tex
\begin{figure}[t]\centering
\begin{tikzpicture}
\begin{axis}[width=0.94\columnwidth, height=5.6cm, x dir=reverse,
  xlabel={emergency-rating uplift $r^C/r^A$ (tightening $\rightarrow$)},
  ylabel={\rev{constraints}}, ymin=0, ymax=14, xmin=0.98, xmax=1.52,
  xtick={1.5,1.4,1.3,1.2,1.1,1.0},
  tick label style={font=\scriptsize}, label style={font=\small},
  legend style={font=\scriptsize, at={(0.5,-0.30)}, anchor=north, legend columns=2, cells={anchor=west}, /tikz/every even column/.append style={column sep=6pt}}, grid=major,
  axis y line*=left]
\addplot[thick, mark=*, mark size=1.5pt] coordinates {(1.5,8) (1.4,12) (1.3,11)};
\addlegendentry{case118: $|U|$}
\addplot[thick, dashed, mark=o, mark size=1.5pt] coordinates {(1.5,6.26) (1.4,8.53) (1.3,8.78)};
\addlegendentry{case118: $\mathbb E|S|$}
\addplot[thick, mark=square*, mark size=1.4pt] coordinates {(1.5,4) (1.4,6) (1.3,6) (1.2,7) (1.15,9) (1.1,10) (1.05,12) (1.02,9) (1.0,10)};
\addlegendentry{case500\_goc: $|U|$}
\addplot[thick, dashed, mark=square, mark size=1.4pt] coordinates {(1.5,2.58) (1.4,2.61) (1.3,3.06) (1.2,2.86) (1.15,2.93) (1.1,3.73) (1.05,4.56) (1.02,5.33) (1.0,6.0)};
\addlegendentry{case500\_goc: $\mathbb E|S|$}
\addlegendimage{thick, dotted, mark=triangle*, mark size=1.7pt}
\addlegendentry{case118: feasible fraction}
\addlegendimage{thick, dotted, mark=triangle, mark size=1.7pt}
\addlegendentry{case500\_goc: feasible fraction}
\end{axis}
\begin{axis}[width=0.94\columnwidth, height=5.6cm, x dir=reverse,
  xmin=0.98, xmax=1.52, ymin=0, ymax=1.05, axis y line*=right, axis x line=none,
  ylabel={feasible fraction}, ylabel style={font=\small},
  tick label style={font=\scriptsize}]
\addplot[thick, dotted, mark=triangle*, mark size=1.7pt] coordinates {(1.5,1.0) (1.4,0.92) (1.3,0.18) (1.2,0.0)};
\addplot[thick, dotted, mark=triangle, mark size=1.7pt] coordinates {(1.5,1.0) (1.4,1.0) (1.3,1.0) (1.2,0.95) (1.15,0.88) (1.1,0.44) (1.05,0.16) (1.02,0.06) (1.0,0.03)};
\end{axis}
\end{tikzpicture}
\caption{Tightening the emergency rating at fixed load. Left axis: \rev{umbrella size} $|U|$ (solid) and per-instance active set $\mathbb E|S|$ (dashed). Right axis: fraction of the $100$ instances that remain feasible (dotted). Case118's window closes within one rung with $\rho\ge0.71$; case500's stays open longer and $\rho$ dips to $0.33$, but $|U|$ never exceeds $12$ \rev{constraints}.}
\label{fig:tightness}
\end{figure}
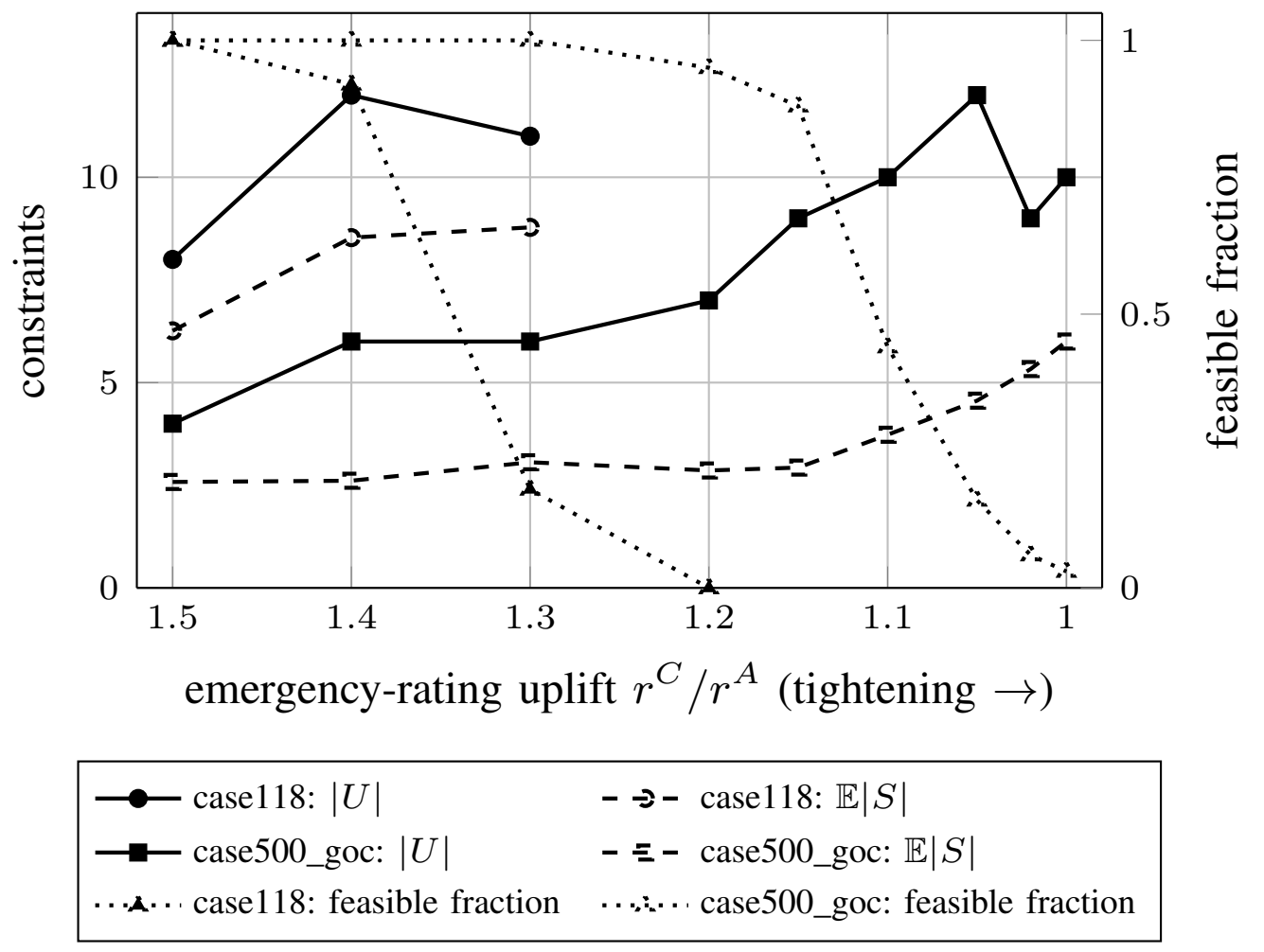

%% file: tables/tab_falsify.tex
\begin{table*}[t]
\caption{Falsification attempts on a fixed network. $\rho=\E|S|/|U|$. None opens a regime for learned selection.}
\label{tab:falsify}
\centering\footnotesize\setlength{\tabcolsep}{4pt}
\begin{tabular}{lcccl}
\toprule
Lever & $|U|$ & $\E|S|$ & $\rho$ & Outcome \\
\midrule
Grid size: case500\_goc (pre-registered) & $6$ & $2.61$ & $0.44$ & \rev{umbrella} \emph{shrank}; hypothesis falsified \\
Load spread: $\pm40\%$ envelope, case118 & $16$ & $6.18$ & $0.39$ & umbrella exactly feasible at every load scale \rev{up to $1.1\times$}; selectors degrade \\
Contingency order: full N-2, case118 & $13$ & $8.3$ & $0.64$ & $327$k candidates; saturates by $n\approx30$ \\
Corrective SCOPF, case2000\_goc & $161$ & $54$ & $0.34$ & ramp \rev{constraints} scale, but \rev{umbrella} $=$ per-instance max \\
Library sweep, \rev{43} networks ($3$--$9\,591$ buses) & med.\ $5$ & --- & med.\ $\rev{0.67}$ & max $\rev{279}$ at $\rho=\rev{0.58}$; no size trend ($r=\rev{0.15}$) \\
Rating tightness: $r^C$ to the boundary & $\le12$ & $\le6$ & $0.33$--$0.80$ & $|U|$ stays tens; window closes by infeasibility \\
\bottomrule
\end{tabular}
\end{table*}

%% file: figures/fig_library.tex
\begin{figure}[t]\centering
\begin{tikzpicture}
\begin{axis}[width=0.98\columnwidth, height=5.4cm, xmode=log,
  xlabel={network size (buses)}, ylabel={\rev{umbrella size} $|U|$},
  ymin=0, ymax=290, xmin=2.5, xmax=12000,
  tick label style={font=\scriptsize}, label style={font=\small}, grid=major]
\addplot[only marks, mark=*, mark size=1.8pt, revcolor] coordinates {
 (3,0) (5,1) (14,1) (30,5) (30,1) (39,5) (57,1) (60,17) (89,1) (118,8)
 (179,39) (197,2) (200,0) (240,35) (300,12) (500,5) (588,22) (793,8)
 (1354,4) (1888,13) (1951,5) (2000,5) (2312,21) (2742,1) (2848,1)
 (2868,14) (2869,44) (3022,128) (3970,3) (4020,0) (4601,1) (4619,0)
 (4661,57) (4837,19) (4917,279) (5658,0) (6468,16) (6470,4) (6495,12)
 (6515,32) (7336,3) (9241,25) (9591,0)};
\node[font=\scriptsize, anchor=east] at (axis cs:4917,279) {case4917\_goc, $\rho{=}\rev{0.58}$~};
\node[font=\scriptsize, anchor=east] at (axis cs:3022,128) {case3022\_goc~};
\end{axis}
\end{tikzpicture}
\caption{\rev{Umbrella size} across the PGLib-OPF library ($\rev{43}$ measured networks, $200$ instances each, most congested securely servable operating point).}
\label{fig:library}
\end{figure}
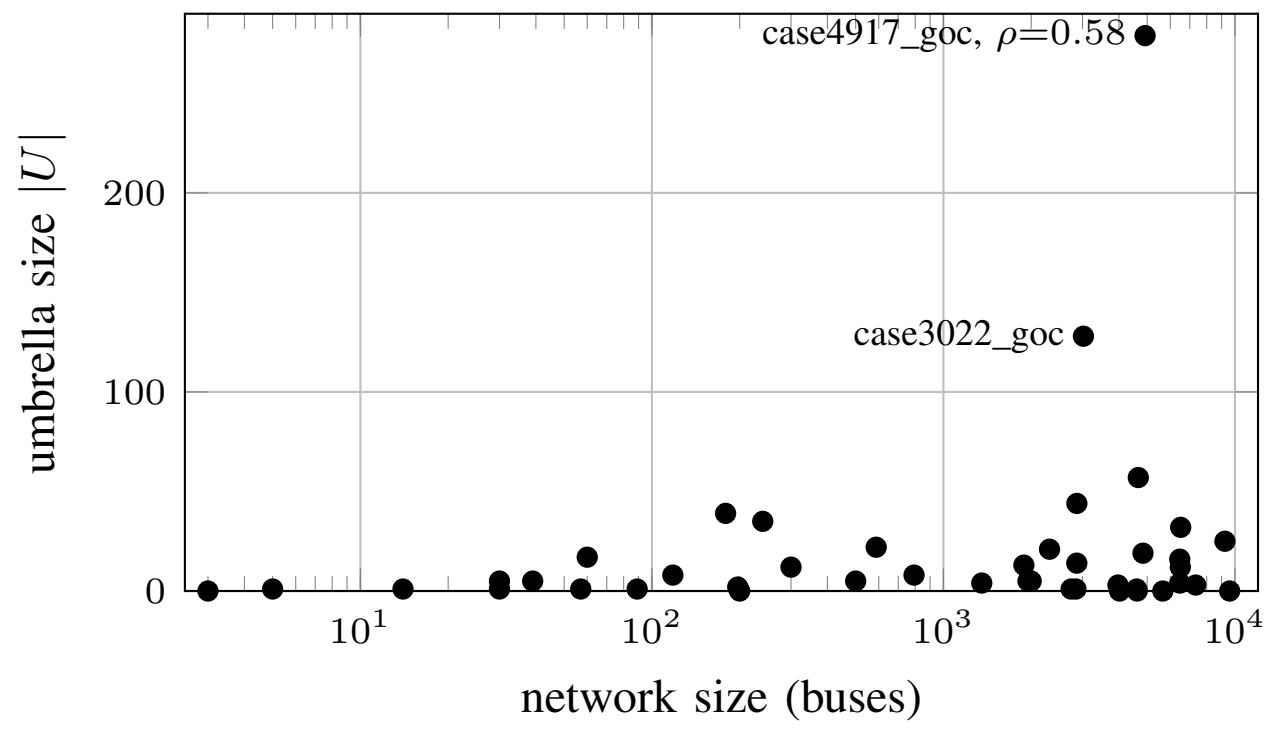

%% file: sections/06_reconcile.tex
\section{Reconciling with the Published Speed-ups}
\label{sec:reconcile}

\rev{The L2O literature reports} computational acceleration \rev{of} two to four orders of magnitude from learned selection \rev{\cite{misra2022learning,deka2019learning,le2026hybrid}. These are online, per-query figures that exclude the offline cost of labelling and training, measured against solving the full problem}. However, both these results and those of our paper are correct \rev{albeit} they use different baselines. Our own pipeline reproduces \rev{speed-ups of this magnitude} before losing to the umbrella. Against our exact solver (Clarabel \rev{over the full screened constraint set}), the selector posts the advertised result at case500: \rev{about $173$}\,s to $\rev{2.15}$\,ms (\rev{about $\times80\,000$} median), binding-set recall $\rev{0.984}$, F$_2=\rev{0.980}$, near-zero optimality gap\rev{ (relative cost difference from the exact optimum)}. Re-baselined against the umbrella on the same test set \rev{and in the same timing jobs}, the umbrella answers in $\rev{2.22}$\,ms to the selector's $\rev{2.15}$\,ms, is \rev{exactly feasible ($0\%$ of instances with any violation, against $4.6\%$ for the selector)} and needs no restoration. \rev{The speed-up therefore belongs to the reduced solve rather than to the learned selection, and the umbrella collects nearly all of it (about $\times78\,000$).} The selector's true margin is \rev{$0.07$\,ms per query ($0.23$\,ms in a paired, interleaved measurement; $0.03$\,ms at case118), and it is} bought with a violation rate. \rev{The selector is also barely per-instance: over the $3\,000$ test queries the hybrid predicts only $8$ distinct constraint sets at case118 and $6$ at case500, and SSL at case500 predicts one $12$-constraint set on $2\,999$ of them.}

Amortisation does not change this. On case118 the \rev{umbrella} is within one \rev{constraint} of its final \rev{size} after $25$ labelled instances and \rev{reaches it} by $2\,000$, so the umbrella's offline cost ($750$\,s of label solves) is a by-product of labels a supervised selector needs anyway, while selector training consumed $10\,000$ labels ($5\times$ the solves) plus training time. To repay the difference the selector would need to save $3.0$\,s per query over $1\,000$ queries or $30$\,ms over $100\,000$, against an available saving of \rev{about $0.03$\,ms per query at case118 (Table~\ref{tab:h2h}), so the account balances only after about $10^8$ queries, and only if the selector's violations are ignored}. ``Learned selection beats solving everything'' is true and was never in doubt. However the deployable question is ``learned selection beats the best trivial reduced solve''\rev{,} and on fixed networks our results \rev{indicate that it does not}. 

%% file: sections/07_topology.tex
\section{Configuration Diversity and Certified Coverage}
\label{sec:topology}
Every result above assumes one fixed network. Operators run a different configuration every week (maintenance outages, unit unavailability, switching), and the umbrella they maintain is the union over configurations. This is the one lever that moves, and it moves the coverage channel (C3).

\subsection{Design}
OPFData's \cite{opfdata2024} component-outage release supplies the same load distribution over networks with one component out of service. Two undocumented properties shaped the design\rev{:} the dataset contains no intact network, \rev{so} a reference network is reconstructed as the union of components ever seen, and the removed component is sometimes a generator ($52$ of $224$ case118 configurations), \rev{so instances must be grouped into configurations by both the branch and the generator that is out of service} (grouping \rev{by} branches alone solved \rev{generator-outage} instances \rev{with the wrong generator limits} and produced spurious infeasibility on about a third of them). \rev{Each constraint is matched across configurations by an identifier built from its monitored and outaged branches, each identified by its terminal buses, reactance and a parallel-circuit ordinal; we verified that this identifier is one-to-one on the screened constraints of the intact case118 and case500 networks.} The \rev{security criterion} is N-1 on a network already one component down.

\subsection{Result and scale check}
Table~\ref{tab:topo} gives the result on three networks. Within any single configuration nothing changes: case118's per-configuration \rev{umbrella has} $8.43\pm0.86$ \rev{constraints}, reproducing the fixed-network $9$, with $\rho_t=0.75$. Pooled, however, its umbrella \rev{has} $57$ \rev{constraints,} of which a typical instance binds one in nine, and the growth curve over configurations does not flatten ($|U(T)|=8.5,13.4,27.4,38.3,47.2,54.1,57.0$ at $T=1,10,50,100,150,200,221$); a Chao2 richness estimate \cite{chao1987} places the asymptote near $97$ (Fig.~\ref{fig:growth}). The composition is a small core plus a long tail in that only $8$ \rev{constraints} bind in a quarter or more of configurations, and $48$ of the $57$ never bind on the intact network across $15\,000$ instances. This is a qualitatively different failure of the static policy \rev{from the one in unit commitment (Section~\ref{sec:uc}). There, a static commitment mostly remains feasible but pays for spare units, so the failure is one of cost; here, a fresh configuration can bind a constraint that no previously seen configuration has bound, and which is therefore absent from the umbrella, so the failure is one of safety.} The effect weakens monotonically with scale and has vanished by $2\,000$ buses, where the pooled \rev{umbrella} equals the within-configuration \rev{umbrella}, every \rev{constraint} is shared by essentially every configuration, and all $120$ configurations solve $20/20$ instances feasibly.

\rev{In the OPFData case2000 release, the $120$ best-sampled configurations (each with at least $25$ instances) are all} generator \rev{ outages}, whereas case118's \rev{coverage failure} was driven mostly by branch outages ($172$ of $224$). We therefore \rev{tested branch outages separately} with a synthetic branch-outage sweep on the PGLib network: one non-islanding branch removed at a time, outages chosen adversarially (the corridors the intact \rev{umbrella} itself monitors) as well as at random. Over $101$ such configurations the pooled \rev{umbrella has} $10$ \rev{constraints}, a $4$-\rev{constraint} core binding in $100$ of $101$ configurations plus a $6$-\rev{constraint} tail with each tail \rev{constraint} observed in at most $5$ configurations, feasibility $18$--$20/20$ everywhere. Branch outages do produce a case118-type tail where generator outages produced none, and it creeps rather than saturates ($|U(T)|=4.0,6.4,7.8,10.0$ at $T=1,25,50,101$), but at an order of magnitude smaller amplitude and with no feasibility damage. A cheap diagnostic, the median number of configurations in which a pooled \rev{constraint} binds, separates the two \rev{behaviours}: below about $200$ buses in the library the pooled/within ratio runs $2.2$--$4.5$ with the median \rev{constraint} shared by $8$--$23\%$ of configurations, whereas from $240$ buses upward the ratio is $1.0$--$1.5$ and the median \rev{constraint} is shared by $85$--$100\%$. On the joint reading (low sharing \emph{and} a large pooled \rev{umbrella}) no library network above $200$ buses appears at risk. We would state the claim as: configuration diversity \emph{can} break static coverage, and case118 shows the regime exists, not that it does so at scale.

\input{tables/tab_topo}

\input{figures/fig_growth}

\subsection{Certifying coverage}
\rev{Once the umbrella can no longer be enumerated, its coverage has to be certified rather than assumed.} In this section we introduce a statistical certificate \rev{that bounds the probability that a fresh configuration binds a constraint outside a given enforced set}. 

 We order candidate \rev{constraints} by the number of \emph{family} configurations in which they bind and let $\widehat U_\lambda$ be the top $\lambda$, a nested family. We then define the miss loss $\ell_\lambda(c)=\mathbf 1[S(c)\not\subseteq\widehat U_\lambda]$, monotone non-increasing in $\lambda$ and use conformal risk control \cite{angelopoulos2024crc} on a separate calibration split of $n$ configurations to pick $\hat\lambda=\inf\{\lambda:\tfrac{n}{n+1}\widehat R_n(\lambda)+\tfrac1{n+1}\le\alpha\}$, which gives $\E[\ell_{\hat\lambda}(c_{n+1})]\le\alpha$ over the joint draw, checked on a held-out test split. The guarantee is operator-readable: enforce these $\hat\lambda$ constraints and a fresh configuration binds something unenforced with probability at most $\alpha$. The three-way split is not optional; building the ordering from the calibration configurations makes the family data-dependent, voids the guarantee and measurably violated it ($16.4\%$ empirical miss at $\alpha=0.05$). Only family-observed \rev{constraints} are enforceable, and a \rev{constraint} first seen in a test configuration counts as a miss.

The bottom of Table~\ref{tab:topo} shows \rev{how the certificate assesses the coverage of the umbrella on the three networks}. On case118 no enforced set of any size certifies $\alpha\le0.05$: even enforcing every \rev{constraint} observed across the family configurations misses $16.9\%$ of fresh ones. On case500 five \rev{constraints} certify $\alpha=0.05$ on $78\%$ of splits and seven \rev{constraints} miss $2\%$. On case2000 the full $8$-\rev{constraint umbrella} certifies every tested level with zero empirical misses across $500$ splits ($\alpha=0.01$ is unreachable only because about $40$ calibration configurations bound CRC at $1/(n+1)\approx0.024$ even at zero risk). The certificate distinguishes the network where static enforcement is safe from the one where it is not, from held-out data, with no assumption beyond exchangeability of configurations.

%% file: tables/tab_topo.tex
\begin{table*}[t]
\caption{Configuration diversity and certified coverage. CRC rows give the fraction of $500$ \rev{($300$ for case118)} family/calibration/test splits on which level $\alpha$ is achievable and the mean test miss rate; a level achievable on only a sliver of splits (case118, $\alpha=0.10$) is conditioned on a rare calibration event and should not be read as reliably achievable.}
\label{tab:topo}
\centering\footnotesize\setlength{\tabcolsep}{4pt}
\begin{tabular}{lccc}
\toprule
 & case118 & case500\_goc & case2000\_goc \\
\midrule
configurations measured & $221$ (\rev{$169$} branch, $52$ gen.) & $150$ ($30$ branch, $120$ gen.) & $120$ (gen.; $+101$ synthetic branch) \\
within-configuration $|U_t|$ & $8.43\pm0.86$ & $4.01\pm0.57$ & $7.82\pm0.43$ \\
pooled $|U|$ (pooled / within) & $57$ ($6.8\times$) & $9$ ($2.2\times$) & $8$ ($1.02\times$); $10$ under branch outages \\
$\rho$ pooled & $0.11$ & $0.27$ & $0.87$ \\
median configurations per pooled \rev{constraint} & $2/221$ & $14/150$ & $119/120$ \\
\midrule
CRC $\alpha=0.20$: achievable / test miss & $74\%$ / $0.202$ & $100\%$ / $0.096$ & $100\%$ / $0.000$ \\
CRC $\alpha=0.10$ & $3\%$ / $0.182$ & $99\%$ / $0.061$ & $100\%$ / $0.000$ \\
CRC $\alpha=0.05$ & never & $78\%$ / $0.031$ & $100\%$ / $0.000$ \\
enforce all family-observed \rev{constraints}: \rev{count} / test miss & $33$ / $0.169$ & $7$ / $0.020$ & $8$ / $0.000$ \\
\bottomrule
\end{tabular}
\end{table*}

%% file: figures/fig_growth.tex
\begin{figure}[t]\centering
\begin{tikzpicture}
\begin{axis}[width=0.98\columnwidth, height=5.4cm, xmode=log,
  xlabel={configurations pooled, $T$}, ylabel={pooled \rev{umbrella} $|U(T)|$}, ymin=0, ymax=62,
  tick label style={font=\scriptsize}, label style={font=\small},
  legend style={font=\scriptsize, at={(0.02,0.98)}, anchor=north west, cells={anchor=west}}, grid=major]
\addplot[thick, mark=*, mark size=1.5pt] coordinates {(1,8.5) (10,13.4) (50,27.4) (100,38.3) (150,47.2) (200,54.1) (221,57.0)};
\addlegendentry{case118, OPFData ($221$)}
\addplot[thick, mark=triangle*, mark size=1.8pt] coordinates {(1,4.05) (2,4.27) (5,4.68) (10,5.2) (25,5.85) (50,6.87) (100,7.87) (150,9.0)};
\addlegendentry{case500\_goc, OPFData ($150$)}
\addplot[thick, mark=square*, mark size=1.4pt] coordinates {(1,7.82) (2,7.96) (5,8.0) (10,8.0) (25,8.0) (50,8.0) (100,8.0) (120,8.0)};
\addlegendentry{case2000\_goc, OPFData gen.\ outages ($120$)}
\addplot[thick, dashed, mark=diamond*, mark size=1.8pt] coordinates {(1,4.05) (5,4.46) (10,5.04) (25,6.35) (50,7.82) (75,8.96) (101,10.0)};
\addlegendentry{case2000\_goc, synthetic branch outages ($101$)}
\end{axis}
\end{tikzpicture}
\caption{\rev{Size of the pooled umbrella} against the number of configurations pooled. Case118 does not saturate; case500 slows; case2000 is flat under generator outages and creeps by six \rev{constraints} under $101$ adversarial and random branch outages.}
\label{fig:growth}
\end{figure}
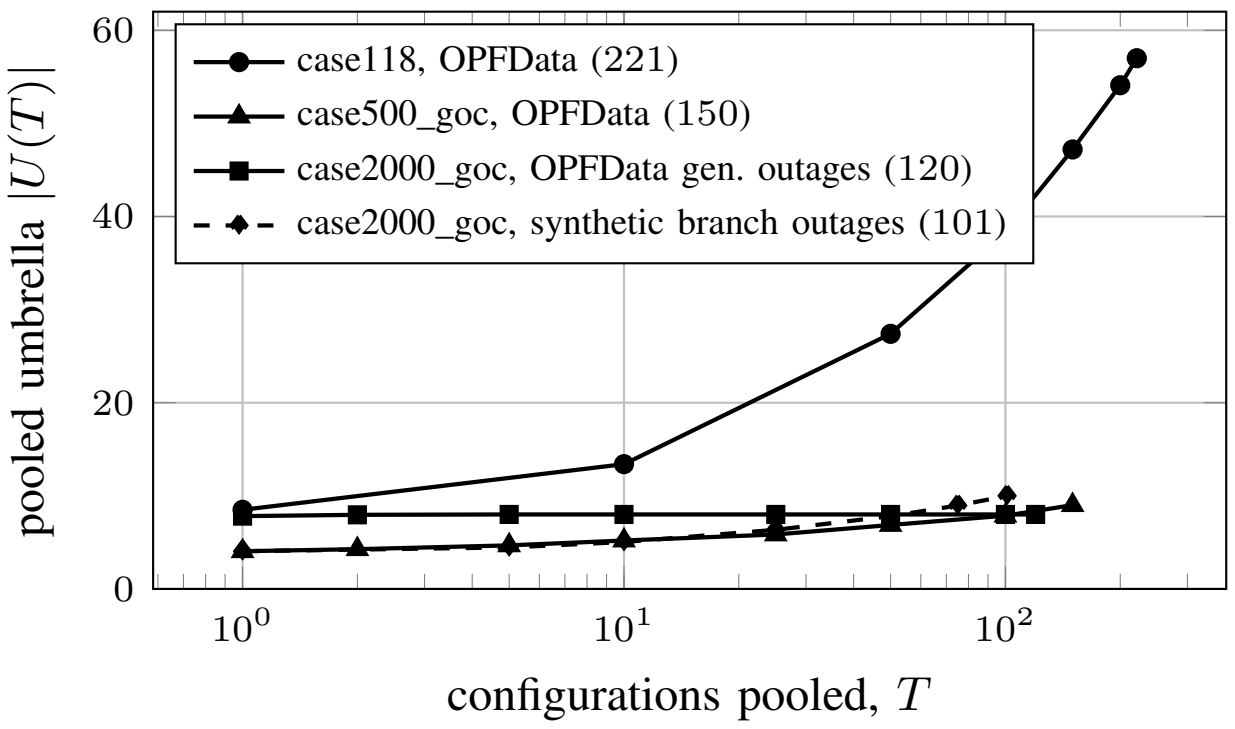

%% file: sections/08_uc.tex
\section{Unit Commitment: \rev{Where a Spare Has a Price}}
\label{sec:uc}
Commitment binaries break the free-spare \rev{property:} a spare committed unit, unlike a spare constraint, costs no-load whether it runs or not, \rev{so} a static commitment is no longer free and \eqref{eq:criterion} predicts a nonzero static gap that grows with the price of the spare. This is \rev{a} regime \rev{for learned selection: a static policy, although it mostly remains feasible, is no longer optimal}. 

\subsection{Model and static gap}
We solve security-constrained unit commitment (SCUC) as a MILP with HiGHS \cite{highs2018}: $\min\ c_1^{\top}\pg+\bm{nl}^{\top}\bm u$ subject to $\bm 1^{\top}\pg=\bm 1^{\top}\dload$, $0\le p_g\le p_g^{\max}u_g$, $F\pg\le h$ (the same N-1 \rev{constraints}), $\bm u\in\{0,1\}^{n_g}$, with no-load $nl_g=\alpha c_{1,g}p_g^{\max}$; case118 has $19$ committable units. \rev{We consider two static commitments fitted on training loads: the \emph{static union}, which commits every unit committed in any training-optimal solution, and the} \emph{best-feasible static commitment}\rev{, obtained by greedily removing units from the union while every training load stays feasible and the total training cost falls. We evaluate both on held-out loads} against instance-optimal SCUC\rev{, against a nearest-neighbour predictor (1-NN) that copies the commitment of the nearest training load, and against the K-candidate method of Pineda and Morales \cite{pineda2022lowhanging} with $K=50$}. Table~\ref{tab:uc} (top) gives the result. The static gap is real and grows with the price of a spare: $+3.7\%$ at $\alpha=0.4$, $+6.9\%$ at $\alpha=0.8$, $+12\%$ under the $\pm40\%$ envelope. \rev{The best-feasible} static commitment always generalises for feasibility in the single-period setting ($0/150$ infeasible)\rev{, so in unit commitment the} static \rev{policy is} suboptimal rather than unsafe. \rev{Nearest-neighbour prediction with a single candidate is} cost-accurate but not feasibility-safe: \rev{1-NN} is within $1.2$--$1.6\%$ of optimal on cost and infeasible on $38$--$43\%$ of held-out loads. \rev{With $K=50$ candidates, however, the method of \cite{pineda2022lowhanging} is feasible on every held-out load and within $0.05$--$0.07\%$ of optimal, confirming their observation in the single-period setting; since this MILP solves in about $0.3$\,s, the benefit of learning there is cost against the static policy rather than speed against the exact solve.} Multi-period commitment ($T=24$, min-up $3$\,h, start-up costs, $30$ train / $20$ test daily profiles) roughly triples the gap to $9.6\%$ at a $1\%$ MILP tolerance (it was $9.3\%$ at $5\%$; tightening the tolerance moves it the direction it must) and makes the best-feasible static schedule infeasible on $5\%$ of unseen days; a fleet-scaling sweep ($1\times$ to $12\times$ the committable units) leaves the multi-period gap at $7$--$9\%$, so it does not grow with fleet size. All headline figures are split-robust: over three random splits the single-period gap is $3.7/2.3/3.7\%$ ($\alpha=0.4$) and $6.9/7.2/6.9\%$ ($\alpha=0.8$) with $0\%$ static infeasibility at every seed, and the multi-period gap is $9.6/10.3/7.9\%$ with $5/10/20\%$ static infeasibility, noisier as $20$ test days must be, with \rev{1-NN} at $25$--$60\%$ infeasible. The identical multi-period pilot at case500 gives $9.7\%$ ($1/20$ infeasible; \rev{1-NN} $0.8\%$ at $25\%$), so the multi-period static gap is scale-stable.

\subsection{The pipeline}
We close the loop by building the \rev{predict-repair-reduce pipeline of Fig.~\ref{fig:pipeline}(b)}. A per-hour neural predictor maps each hour's load features to a commitment, a recovery step commits the cheapest spare units wherever a period cannot be served under security, and a per-period dispatch LP \rev{computes the dispatch and its cost for the resulting commitment}, with no integer variables at deployment. Table~\ref{tab:uc} (bottom) reports out-of-sample results at case118 ($127$ train / $43$ test daily profiles) and at case500 ($171$ dispatchable units, $90/30$)\rev{, against MILP labels solved to a $5\%$ optimality gap}. At case118 the pipeline reaches within $2.9\%$ of the MILP, is feasible on every test day, and runs $161\times$ faster ($60$\,ms vs.\ $9.6$\,s) at $7.4$ recovery commitments per day. Consequently, it dominates the two degenerate pieces from which it has been built, since the bare predictor is $95\%$ infeasible and the best static schedule about three times costlier ($8.6\%$). At case500, nine times the committable fleet, every margin moves the way the mechanism predicts: cost gap $0.2\%$, infeasibility $0/30$, the bare predictor infeasible on $30/30$ days, and the best static schedule $10.1\%$ costlier and itself infeasible on $2/30$ days. \rev{The K-candidate method of \cite{pineda2022lowhanging} improves with $K$ but plateaus at $3/43$ (case118) and $5/30$ (case500) infeasible days from $K=25$ onward, and even using every training day leaves the same days unserved: no historical schedule serves them, and only a repair that creates new commitments can. On the days it does serve, its cost gap is $0.23\%$ at case118 and $1.36\%$ at case500, both within the label tolerance, at $6$ and $51$\,ms per day with the candidates priced in parallel. Our synthetic profiles contain only eight distinct daily shapes, which favours nearest neighbours, so this comparison is conservative for the pipeline.} The speed-up is $16\times$ ($1.07$\,s vs.\ $17.3$\,s), smaller than at case118 because HiGHS solves this MILP well at the $5\%$ tolerance; the pipeline's margin at scale is therefore carried by what the \rev{alternatives lose: a scale-stable cost gap of about $10\%$ for the static schedule and, for nearest-neighbour candidates, the days without precedent, both} of which the pipeline captures essentially \rev{in full}.

\input{tables/tab_uc}

%% file: tables/tab_uc.tex
\begin{table*}[t]
\caption{Unit commitment, out of sample. Top: static commitment gap on case118 (single-period $300/150$; multi-period $30/20$ daily profiles). Bottom: the predict-repair-reduce pipeline. \rev{P\&M: the K-candidate method of \cite{pineda2022lowhanging}, not run on the $30/20$ multi-period splits (see the bottom block). P\&M times are per day with the $K$ candidates priced in parallel.}}
\label{tab:uc}
\centering\scriptsize\setlength{\tabcolsep}{3pt}
\begin{tabular}{lccccc}
\toprule
Setting & static union gap & best-feasible static gap & static test-infeasible & \rev{1-NN} gap (infeas.) & \rev{P\&M, $K{=}50$} \\
\midrule
single-period, $\alpha=0.4$ & $13.96\%$ & $+3.73\%$ & $0/150$ & $1.2\%$ ($38\%$) & \rev{$0.05\%$ ($0\%$)} \\
single-period, $\alpha=0.8$ & $23.08\%$ & $+6.90\%$ & $0/150$ & $1.6\%$ ($41\%$) & \rev{$0.05\%$ ($0\%$)} \\
single-period, $\pm40\%$ envelope & $13.59\%$ & $+12.00\%$ & $0/143$ & $1.3\%$ ($43\%$) & \rev{$0.07\%$ ($0\%$)} \\
multi-period, $1\%$ MILP gap & $22.7\%$ & $+9.6\%$ & $1/20$ & $0.8\%$ ($25\%$) & \rev{---} \\
multi-period, case500\_goc ($5\%$ MILP gap) & $13.2\%$ & $+9.7\%$ & $1/20$ & $0.8\%$ ($25\%$) & \rev{---} \\
\midrule
Pipeline (multi-period) & L2O cost gap (infeas.) & bare predictor infeas. & best static schedule & speed-up vs.\ MILP & \rev{P\&M, $K{=}50$} \\
\midrule
case118 ($127/43$) & $2.9\%$ ($0/43$) & $95\%$ & $8.6\%$ & $161\times$ ($60$\,ms vs.\ $9.6$\,s) & \rev{$0.23\%$ ($3/43$); $6$\,ms} \\
case500\_goc ($90/30$) & $\bm{0.2\%}$ ($\bm{0/30}$) & $100\%$ & $10.1\%$, $2/30$ infeasible & $16\times$ ($1.07$\,s vs.\ $17.3$\,s) & \rev{$1.36\%$ ($5/30$); $51$\,ms} \\
\bottomrule
\end{tabular}
\end{table*}

%% file: sections/09_discussion.tex
\section{Discussion}
\label{sec:discussion}
\subsubsection*{A boundary map rather than a verdict} \rev{One criterion predicts where learned selection is beneficial, and every experiment on both sides of the boundary is consistent with it, which seems to us a more useful claim than either ``L2O works'' or ``L2O does not''. On a fixed network, the reason learned selection does not pay is mechanistic: the umbrellas are small, highly overlapping across instances and quickly saturating, and they remain so under stress-testing. We would stress two further points. First, the negative half does not depend on the quality of our selector: Proposition~\ref{prop:spare} and \eqref{eq:criterion} bound what any selector can deliver, including an oracle that always predicts the true active set, and the measured $|\Uset|$, $\E|S|$ and $T(\cdot)$ make that bound less than a millisecond, so a better model cannot change it. Second, where a spare carries a price, the same learning machinery is both valuable and safe, its margin at scale is carried by what the alternatives lose, and repair is what makes it deployable: prediction without repair fails in both the single- and multi-period settings, for our learned predictors and for a single nearest neighbour, and with $K$ candidates nearest-neighbour prediction still cannot serve the multi-period days that no past commitment covers.}

\subsubsection*{Limitations} The study is DC. A companion AC audit indicates that reduced N-1 restoration built from DC-derived selections is as secure as full restoration at about $17\times$ lower cost, but that DC labels do not deliver the recall an AC-native selector would need. We will address an AC-native selection in a future work. The unit-commitment model is stylised (fixed-charge units with $p^{\min}=0$, DC security, synthetic daily profiles from a canonical shape and per-day amplitude\rev{, with only eight distinct shapes}), and the pipeline results \rev{at both sizes} use a $5\%$ MILP tolerance for labels\rev{; the} case118 tolerance check suggests this understates rather than overstates the static gap. The library sweep evaluates each network at a stressed operating point, and its $\rho$ values are boundary values, and $13$ PGLib networks \rev{could not be made N-1 feasible on our rating--load ladder}. The configuration result at $2\,000$ buses rests on generator outages from OPFData plus synthetic branch outages. \rev{Multi-component outage states, seasonal ratings and switching would widen the configuration distribution over which the certificate is calibrated and would test its one assumption, exchangeability across configurations, which we will address in future work.} 

\subsubsection*{Practical reading} For an operator, the criterion is cheap to apply before committing to a learned pipeline: measure $|\Uset|$, $\E|S|$ and $T(\cdot)$ on the problem at hand and ask whether any spare carries a price in the objective. In cases that it does not, simply enforce the umbrella and certify its coverage. On the other hand, in cases where any spare carries a price, the predict-repair-reduce architecture is the appropriate pipeline to use. Code, labelled datasets and every result file behind the tables will be made available on publication.

%% file: sections/10_conclusion.tex
\section{Conclusion}
\label{sec:conclusion}
\rev{We asked whether learned constraint selection beats enforcing the umbrella, the strongest learning-free alternative, in security-constrained optimisation. On a fixed network we find that it does not, across grid size, load spread, contingency order, corrective redispatch, rating tightness and the whole PGLib-OPF library, and the published speed-ups belong to the reduced solve rather than to the learned selection. Configuration diversity can break the umbrella's coverage on a small, densely meshed network, which conformal risk control detects from held-out data, but the effect is absent at 2\,000 buses. In unit commitment, where a spare has a price, a static commitment loses $4$--$12\%$ and a predict-repair-reduce pipeline recovers it at both fleet sizes tested, including on days that no past commitment can serve. On this evidence, learning pays where a spare has a price.}

%% file: main.bbl
\begin{thebibliography}{10}
\providecommand{\url}[1]{#1}
\csname url@samestyle\endcsname
\providecommand{\newblock}{\relax}
\providecommand{\bibinfo}[2]{#2}
\providecommand{\BIBentrySTDinterwordspacing}{\spaceskip=0pt\relax}
\providecommand{\BIBentryALTinterwordstretchfactor}{4}
\providecommand{\BIBentryALTinterwordspacing}{\spaceskip=\fontdimen2\font plus
\BIBentryALTinterwordstretchfactor\fontdimen3\font minus \fontdimen4\font\relax}
\providecommand{\BIBforeignlanguage}[2]{{%
\expandafter\ifx\csname l@#1\endcsname\relax
\typeout{** WARNING: IEEEtran.bst: No hyphenation pattern has been}%
\typeout{** loaded for the language `#1'. Using the pattern for}%
\typeout{** the default language instead.}%
\else
\language=\csname l@#1\endcsname
\fi
#2}}
\providecommand{\BIBdecl}{\relax}
\BIBdecl

\bibitem{ardakani2013umbrella}
A.~J. Ardakani and F.~Bouffard, ``Identification of umbrella constraints in {DC}-based security-constrained optimal power flow,'' \emph{IEEE Trans. Power Syst.}, vol.~28, no.~4, pp. 3924--3934, 2013.

\bibitem{misra2022learning}
S.~Misra, L.~Roald, and Y.~Ng, ``Learning for constrained optimization: Identifying optimal active constraint sets,'' \emph{INFORMS J. Comput.}, vol.~34, no.~1, pp. 463--480, 2022, arXiv:1802.09639.

\bibitem{deka2019learning}
D.~Deka and S.~Misra, ``Learning for {DC-OPF}: Classifying active sets using neural nets,'' in \emph{IEEE Milan PowerTech}, 2019, arXiv:1902.05607.

\bibitem{le2026hybrid}
V.-A. Le, M.~Xie, and R.~Mangharam, ``A hybrid learning-to-optimize framework for mixed-integer quadratic programming,'' in \emph{Proc. L4DC}, 2026, arXiv:2511.19383.

\bibitem{stott2012opf}
B.~Stott and O.~Alsa\c{c}, ``Optimal power flow: Basic requirements for real-life problems and their solutions,'' \emph{SEPOPE / white paper}, 2012, widely circulated industry white paper on OPF requirements.

\bibitem{ardakani2015acceleration}
A.~J. Ardakani and F.~Bouffard, ``Acceleration of umbrella constraint discovery in generation scheduling problems,'' \emph{IEEE Trans. Power Syst.}, vol.~30, no.~4, pp. 2100--2109, 2015.

\bibitem{ardakani2018prediction}
------, ``Prediction of umbrella constraints,'' in \emph{Proc. PSCC}, 2018.

\bibitem{ng2018statistical}
Y.~Ng, S.~Misra, L.~A. Roald, and S.~Backhaus, ``Statistical learning for {DC} optimal power flow,'' in \emph{Proc. PSCC}, 2018.

\bibitem{robson2019learning}
A.~Robson, M.~Jamei, C.~Ududec, and L.~Mones, ``Learning an optimally reduced formulation of {OPF} through meta-optimization,'' \emph{arXiv:1911.06784}, 2019.

\bibitem{pineda2020data}
S.~Pineda, J.~M. Morales, and A.~Jim{\'e}nez-Cordero, ``Data-driven screening of network constraints for unit commitment,'' \emph{IEEE Trans. Power Syst.}, vol.~35, no.~5, pp. 3695--3705, 2020, arXiv:1907.04694.

\bibitem{porras2023cost}
{\'A}.~Porras, S.~Pineda, J.~M. Morales, and A.~Jim{\'e}nez-Cordero, ``Cost-driven screening of network constraints for the unit commitment problem,'' \emph{IEEE Trans. Power Syst.}, vol.~38, no.~1, pp. 42--51, 2023, arXiv:2104.05746.

\bibitem{wu2024efficient}
X.~Wu \emph{et~al.}, ``Efficient unit commitment constraint screening under uncertainty,'' \emph{arXiv:2408.05185}, 2024.

\bibitem{yang2023costaware}
S.~Yang \emph{et~al.}, ``Cost-aware bound tightening for constraint screening in {AC} {OPF},'' \emph{arXiv:2305.10385}, 2023.

\bibitem{pineda2022lowhanging}
S.~Pineda and J.~M. Morales, ``Is learning for the unit commitment problem a low-hanging fruit?'' \emph{Electr. Power Syst. Res.}, vol. 207, p. 107851, 2022, arXiv:2106.11687.

\bibitem{amos2017optnet}
B.~Amos and J.~Z. Kolter, ``{OptNet}: Differentiable optimization as a layer in neural networks,'' in \emph{Proc. ICML}, 2017.

\bibitem{donti2021dc3}
P.~L. Donti, D.~Rolnick, and J.~Z. Kolter, ``{DC3}: A learning method for optimization with hard constraints,'' in \emph{Proc. ICLR}, 2021.

\bibitem{zhao2023ensuring}
T.~Zhao, X.~Pan, M.~Chen, and S.~H. Low, ``Ensuring {DNN} solution feasibility for optimization problems with convex constraints and its application to {DC} optimal power flow,'' \emph{arXiv:2112.08091}, 2023.

\bibitem{venzke2020worstcase}
A.~Venzke, G.~Qu, S.~Low, and S.~Chatzivasileiadis, ``Learning optimal power flow: Worst-case guarantees for neural networks,'' \emph{IEEE SmartGridComm}, 2020.

\bibitem{nellikkath2022pinn}
R.~Nellikkath and S.~Chatzivasileiadis, ``Physics-informed neural networks for {AC} optimal power flow,'' \emph{Electr. Power Syst. Res.}, vol. 212, p. 108412, 2022.

\bibitem{chen2024scalable}
W.~Chen, M.~Tanneau, and P.~Van~Hentenryck, ``Scalable exact verification of optimization proxies for large-scale optimal power flow,'' \emph{arXiv:2405.05572}, 2024.

\bibitem{vovk2005algorithmic}
V.~Vovk, A.~Gammerman, and G.~Shafer, \emph{Algorithmic Learning in a Random World}.\hskip 1em plus 0.5em minus 0.4em\relax Springer, 2005.

\bibitem{angelopoulos2023gentle}
A.~N. Angelopoulos and S.~Bates, ``Conformal prediction: A gentle introduction,'' \emph{Found. Trends Mach. Learn.}, vol.~16, no.~4, pp. 494--591, 2023.

\bibitem{angelopoulos2024crc}
A.~N. Angelopoulos, S.~Bates, A.~Fisch, L.~Lei, and T.~Schuster, ``Conformal risk control,'' in \emph{Proc. ICLR}, 2024.

\bibitem{angelopoulos2021learn}
A.~N. Angelopoulos, S.~Bates, E.~J. Cand{\`e}s, M.~I. Jordan, and L.~Lei, ``Learn then test: Calibrating predictive algorithms to achieve risk control,'' \emph{arXiv:2110.01052}, 2021.

\bibitem{patel2025conformal}
R.~Patel \emph{et~al.}, ``Conformal mixed-integer constraint learning with feasibility guarantees,'' \emph{arXiv:2506.03531}, 2025.

\bibitem{zhang2025tightening}
Y.~Zhang \emph{et~al.}, ``Tightening optimality gap with confidence through conformal prediction,'' \emph{arXiv:2503.04071}, 2025.

\bibitem{bengio2013estimating}
Y.~Bengio, N.~L{\'e}onard, and A.~Courville, ``Estimating or propagating gradients through stochastic neurons for conditional computation,'' \emph{arXiv:1308.3432}, 2013.

\bibitem{opfdata2024}
S.~Lovett, M.~Zamora, A.~Faddoul, R.~Ross, E.~Kaufmann, M.~de~Andrade, W.~L. Yeo, D.~Bregman, A.~Ni, Z.~Zhou, A.~Ferrone, N.~Puri, C.~Coffrin, A.~Yousefi, Z.~Zhang, A.~Boehm, M.~Wilhelm, E.~Marino, and A.~Sanchez-Gonzalez, ``{OPFData}: Large-scale datasets for {AC} optimal power flow with topological perturbations,'' arXiv:2406.07234, 2024.

\bibitem{clarabel2024}
P.~J. Goulart and Y.~Chen, ``Clarabel: An interior-point solver for conic programs with quadratic objectives,'' arXiv:2405.12762, 2024.

\bibitem{pglib2019}
S.~Babaeinejadsarookolaee, A.~Birchfield, R.~D. Christie, C.~Coffrin, C.~DeMarco, R.~Diao, M.~Ferris, S.~Fliscounakis, S.~Greene, R.~Huang, C.~Josz, R.~Korab, B.~Lesieutre, J.~Maeght, T.~W.~K. Mak, D.~K. Molzahn, T.~J. Overbye, P.~Panciatici, B.~Park, J.~Snodgrass, A.~Tbaileh, P.~Van~Hentenryck, and R.~Zimmerman, ``The power grid library for benchmarking {AC} optimal power flow algorithms,'' arXiv:1908.02788, 2019.

\bibitem{chao1987}
A.~Chao, ``Estimating the population size for capture-recapture data with unequal catchability,'' \emph{Biometrics}, vol.~43, no.~4, pp. 783--791, 1987.

\bibitem{highs2018}
Q.~Huangfu and J.~A.~J. Hall, ``Parallelizing the dual revised simplex method,'' \emph{Math. Program. Comput.}, vol.~10, no.~1, pp. 119--142, 2018.

\end{thebibliography}
